# THE TRIPLE HELIX MODEL AND THE STUDY OF KNOWLEDGE-BASED INNOVATION SYSTEMS

**Loet Leydesdorff**
*University of Amsterdam*

**ABSTRACT**

This paper examines the changing nature of knowledge-based innovation systems in light of the dynamic interconnections between the university, industry and government. Industries have to assess in what way and to what extent they decide to internalize R&D functions. Universities position themselves in markets, both regionally and globally. Governments make informed trade-offs between investments in industrial policies, S&T policies, and/or delicate and balanced interventions at the structural level. Such policies can be expected to be successful insofar as one can anticipate and/or follow trends according to the dynamics of the new technologies in their different phases. The evolutionary perspective in economics can be complemented with a turn towards reflexivity in sociology in order to obtain a richer understanding of how the overlay of communications in university-industry-government relations reshapes the systems of innovations that are currently subjects of debate, policy-making, and scientific study.

**INTRODUCTION**

Various scholars have proposed categories for the analysis of the ongoing changes in research and innovation systems. Gibbons et al. (1994), for example, distinguished between 'Mode 1' and 'Mode 2' in the production of scientific knowledge. 'Mode 1' research can be considered as disciplinarily organized, while 'Mode 2' research is mainly legitimated and organized with reference to contexts of application. Other policy analysts have argued that systems of innovation can no longer be stabilized nationally, since they remain fundamentally in transition (e.g., Cozzens et al.1990). Rip and Van der Meulen (1996) proposed the concept of a 'post-modern research system' to describe S&T policy systems.

What generates this lack of consensus about the appropriate unit of analysis in the study of technology and innovation? National systems of innovation (Lundvall 1988; Nelson 1993) have been particularly studied because of their relevance for government policies (Nelson 1982; Irvine & Martin 1984; Freeman 1987). Or should one consider 'Research, Technology, & Development' (RTD) systems as an increasingly steady state (Ziman 1994)? Sometimes various claims are made within a single text or the contradicting statements are sometimes combined in edited volumes.

In my opinion, the various metaphors in the study of knowledge-based innovation systems can be considered as theoretical appreciations of a complex dynamics from different perspectives and with potentially different objectives. The analysts attempt to stabilize a picture by choosing a perspective. Since the systems under study are developing dynamically, the metaphors remain 'out of focus' when viewed from one window of appreciation or another. The illusion of a stable object, however, may enable policy advisors to legitimate S&T policies.

Innovation is not a stable unit of analysis, since it operates at an interface by definition. One is able to provide reflections on this operation from different angles and/or at different moments in time. The various story lines illustrate the symmetry breaches at the relevant interfaces reflexively. From this meta-perspective, knowledge-based developments can be considered as systems that are discursively reconstructed as reflections on specific subdynamics (e.g., knowledge production) while at the same time developing in interaction with other subsystems of society (e.g., diffusion at the market). The dynamics of such non-linear interactions are non-trivial and nearly incommensurable windows of appreciation can be expected. In this contribution, I foreground the self-organizing potentials of the complex dynamics of innovation and the evolutionary function of the variety of reflections on this phenomenon.

## THE COMPLEX DYNAMICS OF INNOVATION

While each innovation can be considered as an instance of interaction between different subdynamics, innovation *systems* build recursively on series of interaction terms. The recursion in the interaction can be expected to remain partially beyond control—since at the (next-order) network level—when analyzed from the perspective of either of the subsystems that interact. In other words, the model has to contain both interactive and recursive terms at different levels of aggregation and it may operate counter-intuitively.

In such a non-linear model, intentional input can no longer be expected to lead to intended output. The dependency relations themselves can be expected to change when the systems under study are further developing. Unintended consequences (e.g., economic externalities) can be expected to prevail, while 'externalities' emerge in unexpected contexts (Callon 1998). Contexts become crucial when consequences can no longer unambiguously be related to causes (Barnes and Edge 1982). While policy makers may be able to steer the developments in some cases and at some stages, the policy inputs can be expected to remain contextual in other instances—for example, when the systems internalize complexity by becoming increasingly knowledge-intensive (Van den Daele et al. 1979).

These externalities of innovation systems can be made visible only reflexively, that is, *ex post*. From this perspective of hindsight, however, the focus of the analysis shifts from the social construction of technology (e.g., Bijker et al. 1987) to the appreciation of the complex dynamics of innovation within the constructed system. The social construction by agency can then be considered as the subdynamics of producing variation, whereas selection is structured at other levels.

Each subdynamics can be made the subject of theoretical reflections by taking a perspective or making specific assumptions; for example, about the role of agency or structure (Giddens 1984). The theoretical discourses attempt to stabilize a discursive representation of one of the interacting subdynamics. In addition to the theoretical task of improving on the quality of these theoretical reflections, a methodologist is able to raise questions concerning the contingent relations between different forms of appreciative theorizing (Leydesdorff 1997).

The model of evolutionary theorizing in economics, for example, can be recognized as providing a meta-biological perspective in which selection environments are often considered as 'given' for a firm (Andersen 1994). Evolutionary economists have first drawn attention to the non-linear interaction terms between market perspectives and technological options (Mowery and Rosenberg 1979; Freeman 1982). From a sociological perspective, however, neither selection environments nor technological options are biologically given. All interacting subsystems at the social level (technologies, markets, institutions, etc.) have discursively been constructed and are continuously reconstructed.

Nelson (1994) proposed to analyze co-evolutions between technologies and institutions in addition to co-evolutions between markets and technologies. However, institutional selection operates very differently from the dependency relations between technologies and markets (McKelvey 1996). The Triple Helix thesis focuses on the interactions among these various interfaces. How are organizational rigidities among the helices both organized and dissolved? When can these reorganizations be considered as structural adjustments to technological developments (Freeman & Perez 1988)? How does a system of innovations build on stability *and* change? In which phases can change and/or stability be expected to prevail, and why?

The metaphors are needed, since the systems cannot be defined without theorizing. They are not given, but historically constructed. Therefore, they can be further refined by reflections that are stabilized as theoretical discourse at the social level. The codification of these reflections can sometimes be refined when the respective interpretations are disturbed by interactions. Thus, the various metaphors function both as evaluation schemes and as heuristics. The innovation 'system' under study, however, is operational and therefore non-observable. It remains a knowledge-based assumption.

When studying innovation systems one becomes increasingly aware of the dependency of each analytic perspective on the respective definitions. Changes in definitions sometimes provide windows of unexpected opportunities for innovation. Different stakeholders (e.g., academia, industry, government) recombine from their respective perspectives. Recombinations (that is, knowledge-based innovations and reorganizations) may disturb the current discourses to such an extent that new perspectives can be proposed and elaborated.

**EXPECTATIONS, INSTITUTIONS, AND COMMUNICATIONS**

The institutional arrangements in national systems of innovation compete in terms of their respective successes and failures when attempting to grasp the fruits of possible innovations, for example, by trading off changes in their structures (transaction costs) against historical continuities (routines). The various subdynamics operate upon one another without any *a priori* guarantee of harmonization, but under selection pressure. Order can be expected to emerge in one or more directions because of potential resonances between the selecting systems. These 'lock-ins' remain conditioned and constrained by the historical configurations (David 1985; Arthur 1988, 1989, and 1994).

When and where do the emerging conditions fit into each other, and to what extent? The metaphor of an overlay of mutual expectations and exchange relations enables us to analyze these complex dynamics as a result of the interaction among the various subdynamics, while each of the subdynamics can also be expected to operate recursively on the basis of their previous state. Note that the differentiation is continuously reproduced because the innovative integration at an interface is by definition partial. The differentiation is perhaps changed, but then reconstructed so that new rounds of innovative interfacing remain possible.

Furthermore, the analyst may act as a participant (recursively implied) and/or as an external observer. One is additionally able to change one's perspective, for example, when giving normative advice. Corresponding to this double perspective of analyst and participant (Giddens 1976), the emerging overlay can be considered by each actor as subsystemic (that is, as an interface *within* a system) and/or as supersystemic (that is, as a factor in the system's environment). While a supersystemic factor provides a relevant environment for the system of reference, each participant can also be implied in the (re-) construction of the overlay by reflecting on his or her environment. Thus, a double perspective of participant and observer is reflexively reinforced and knowledge-based learning processes are then induced (Leydesdorff 2001a).

From this (neo-) evolutionary perspective, the observable social structures can be considered as successful cases of previous institutionalization and conflict resolution. The structural forces behind the institutionalization and stabilization may remain latent, but they can be hypothesized. In the longer term, institutions can be expected to optimize their relations with relevant environments (for example, by learning to cope with uncertainties). Thus, the knowledge base of these institutions is further developed.

Institutional functionality in a knowledge-based economy also implies reaching across institutional borders on the basis of expectations about how the environments may change when providing opportunities for innovation. For example, industries have to assess in what way and to what extent they decide to internalize R&D functions. Universities position themselves in markets, both regionally and globally. Governments make informed trade-offs between investments in industrial policies, S&T policies, and/or delicate and balanced interventions at the structural level. Such policies can be expected to be successful insofar as one can anticipate and/or follow trends according to the dynamics of the new technologies in their different phases (Freeman and Perez 1988; McKelvey 1996; Giesecke 2000). The management of these interfaces is both an economic imperative and a political challenge, yet knowledge-intensive in the elaboration.

Is this (neo-) evolutionary model a reappraisal of old-fashioned structural-functionalism? In my opinion, the Triple Helix model extends the basis of structural-functionalism by introducing the notion of 'meaning' from symbolic interactionism: social functions are discursively constructed, and they can be deconstructed and reconstructed reflexively. Thus, one can no longer accept a dialectics between ahistorical functions and historical institutions. The functions can be expected to (co-) evolve with the discourses and the institutions. The institutions are needed to carry out the functions, but they can be expected to be changed while doing so. The functions are continuously under reconstruction and the institutional elements of the systems have been generated by these recursive operations.

**THE HISTORICAL RECONSTRUCTION**

This neo-evolutionary model is historically reflexive, since the cultural evolution that it tries to explain, builds on the achievements of the past. Both layers of the system (institutions and functions) have been socially constructed and stabilized, but in different periods of time. First, the communications were functionally differentiated as in the individual revolutions of the 16$^{th}$ and 17$^{th}$ century. The transformations of the 18$^{th}$ century ('modernization') led in the 19$^{th}$ century to the institutional differentiation between the modern state and civic society.

The complex social system builds on the interfaces among both institutions and functions as different mechanisms of differentiation. Institutions can be assessed in terms of their functionality, and functions can be evaluated in terms of their value for the carrying institutions. Functional meanings and institutional meanings can constructively be 'translated' into each other.

Since the interaction terms are based on reflections and therefore not always readily observable, they suppose the declaration of reflexivity as a condition for their discursive reconstruction. This discursive reflexivity can always be made more knowledge-intensive and science-based. Because of the knowledge-intensity of the communications, one is increasingly able to experiment with the interaction terms between structures and functions in the organization of social systems.

I follow the sociological tradition (Marx, Weber, Parsons) in assuming that the functional differentiation of society was constructed during the individual revolutions of the 16$^{th}$ and 17$^{th}$ centuries. Only after the completion of the 'modern' system could institutional implementation of the 'natural' or 'universal' system be legitimated as deliberate reform of social relations ('modernization'). Foucault (e.g., 1972) used the term 'noso-politics' for this reconstruction of society during the Enlightenment of the 18$^{th}$ century.

The institutional differentiation of the state from civic society followed upon the American and French Revolutions and was mainly achieved in the first half of the 19$^{th}$ century. This development has led to a variety of nation states with their respective cultures. From 1870 onwards, a scientific-technological revolution can subsequently be distinguished, gradually shaping a knowledge-based mode of production and distribution at the global level (Braverman 1974).

During this latter process the knowledge production and control functions increasingly ceased to be the exclusive domain of academia (Noble 1977). Functions and institutions could historically be coupled, but there were no longer determinate relationships (Whitley 1984). For example, the American university, to a larger extent than its German predecessor, became also an entrepreneurial locus (Etzkowitz 2001). The position of government changed from that of a principal agent ('King') into that of a controlled bureaucracy negotiating an internal trade-off between facilitating further developments at the level of society and political accountability (e.g., Van den Belt & Rip 1987; cf. Weber 1922).

Table One captures the interactions and differentiations between functions and institutions in a scheme. In the liberal organization of society, science was first considered as a public good, while trade and industrial production were considered private activities. These categories became increasingly interchangeable across institutional interfaces with the further development of the system. Scientific insights could be made useful in industrial practices and industrial (and military) concerns began to guide the heuristics of scientific research programs. These reflexive flexibilities, perhaps rooted in American pragmatism, also influenced the construction of the European Union after W.W. II, since a variety of perspectives must be translated into each other in order to (re-) construct the European dimension (cf. Ronge 1979).

**Table 1**
**The Interaction Between Functional and Institutional Differentiation**
**(Since Approximately 1870)**

|  | *Functions* | |
| --- | --- | --- |
| *Institutions* | *Science* | *Economy* |
| *Public* | Academia; University | Patent legislation; Science, technology, and innovation policies |
| *Private* | Industrial R&D labs; entrepreneurial universities | Trade and Industry |

**THE CULTURAL EVOLUTION OF KNOWLEDGE PRODUCTION**

The new mode of knowledge-based production can be expected to build on the old one(s) as its historical basis. Thus, labels like 'university', 'industry,' and 'government' did not disappear while constructing the transnational system, but they gradually shifted in meaning (Callon and Latour 1981). However, the changes of meaning are not expected to imply a loss of differentiation at the level of the systems under reconstruction. The differentiation hitherto obtained by the social system provides its complex dynamics with the capacity to develop further in response to emerging challenges.

From this perspective, the established systems and their corresponding discourses compete as suboptimal solutions to the problem of organizing and giving meaning to a social world (a 'universe'). One no longer expects a single (optimum) solution based on an undifferentiated integration, a common center or a textbook (as in a high culture; cf. Yamauchi 1986). One expects local suboptima that explore 'hill-climbing' in their relevant environments (Kauffman 1993; Allen 1994). Nation systems, for example, can then be considered as cultures that compete for a share in the economic developments at the global level.

Since social systems remain distributed by their nature, the institutional hill-climbers compete and thereby reshape the distribution of their opportunity structures in relation to one another. Thanks to this reshaping, the landscape itself can be expected to change (Scharnhorst 1998). As in biology, the landscape is rugged with its historical formations. Unlike biology, however, Schumpeter's (1939) 'creative destruction' of existing constructions is part of the reflexive practices of the hill-climbers.

In a dialectical model of social evolution a co-evolution between two subdynamics (e.g., production forces and production relations) can be hypothesized, but a Triple Helix cannot be stabilized or resolved. A model of three helices is sufficiently complex for understanding the complex dynamics of the ongoing transformation processes. The (three) double helices on which the Triple Helix builds, continuously 'lock-in' into local co-evolutions that are expected to 'clot' into provisional solutions shaping the ruggedness of the corresponding landscapes.

The clots of different sizes perform their own 'life'-cycles along historical 'trajectories', for example, at the industry level, while the overall landscape can be considered as a next-order system forming a 'regime' (Dosi 1982). Bifurcations endogenously reshape the levels in series of events. The 'regime' can be considered as a non-linear effect of the trajectories, and the 'trajectories' are the expected consequences of previous lock-ins (Leydesdorff & Van den Besselaar 1998).

In biology, the *rugged fitness landscapes* are 'given' for the various species and provide niches for their survival. In an economy, the niches can be considered as mechanisms for the retention of already adapted environments: how is one able to organize an institutional arrangement so that wealth and jobs can be generated? These social mechanisms, however, can also be deconstructed and reconstructed.

The solution of local conflicts has hitherto been a central function of the nation states and their political economies (Skolnikoff 1993). As the operation of the third helix (knowledge) becomes more pronounced in the re-organization of society, it continuously may destabilize a click between two other helices. Nelson & Winter (1982), for example, noted that technological innovations tend to upset the equilibrating dynamics of the market. The market, however, can only be equilibrated within an institutional setting. Destabilization can thus be considered as an effect of interactions at the regime level (Freeman and Perez 1988).

Technologies may 'click' with state apparatuses into a local fit, like in the energy sector or in health care; industries potentially click with technologies (e.g., David's (1985) QWERTY keyboards and Arthur's (1988) VHS tapes); and industries can click with the state apparatus as in the former Soviet Union. A click excludes a third subdynamic from effectively operating, since the co-evolutionary dynamic can then be considered as temporarily 'locked' (Simon 1969).

For example, the political economies of Eastern Europe were not sufficiently able to make the transition to a knowledge-based economy during the 1970s and 1980s (e.g., Richta et al. 1968). When the Chinese innovation system was confronted with similar problems of integration in the late 1980s, one could reflexively choose for a knowledge-based reform of the political economy (Leydesdorff and Guoping 2001).

A lock-in can have local advantages (e.g., increasing marginal returns) and/or global disadvantages. A 'break out' of a lock-in may open a window on a new market and offer a global (that is, next-order) perspective. Sometimes it provides also a niche for developing a new discipline. ICT (e.g., Nowak and Grantham 2000) and biotechnology (e.g., McKelvey 1996) have been considered as examples of global developments. But the risk of crisis is ever present given the complexity of the dynamics. As another subdynamic increasingly disturbs a local harmonization, the systems can be expected to become 'critical', that is, to drift towards the edges of chaos and potential bifurcation.

**THE PREDICTIVE POWER OF THE TRIPLE HELIX MODEL**

Can a reflexive observer grasp the evolutionary momentum and perform the adjustments in time? What unintended consequences can be made visible *ex ante* using the available reflections? How large are the expected uncertainties? How can the threats be formulated in terms of opportunities?

As noted, the differentiations that have been achieved historically, cannot be dissolved at the system's level without costs. A loss of internal complexity, for example, can be expected to lead to a loss of ability to handle complexity in the relevant environments (e.g., markets). The functional differentiations of knowledge production, wealth creation, and governance operate as feedbacks on institutional task divisions among academia, industry, and government, and *vice versa*. However, one can no longer expect a single or pre-given (e.g., national) order to prevail: the various subdynamics are juxtaposed. Thus, there is always room for improvement, empirical investigation, and change.

When order can be observed, the analyst is able to hypothesize on theoretical grounds how this order was constructed, for example, at the level of nations or sectors. Simulation studies enable us to specify the (sometimes counter-intuitive) expectations given historical configurations and theoretical assumptions about the relevant subdynamics. However, the simulation results require another round of appreciation.

Although an analyst may be able to specify this uncertainty, the assumption of the non-trivial (social) machinery of a Triple Helix with an overlay *adds* to the uncertainty. For example, other players may see options for codification that the analysts could not possibly have seen given their necessarily contingent positions as also participants. Reflexivity and uncertainty prevail in a knowledge-based economy.

Biological evolution theory assumed variation as the driving force and selection to be 'naturally' given. Cultural evolution, however, can be considered as driven and reconstructed both by individuals and groups who make conscious decisions and by the appearance of a variety of unintended consequences of interactions with which one may have to cope in a next stage. Since the sources of innovation in a Triple Helix configuration are not synchronized *a priori*, the possibilities for innovations and rearrangements generate puzzles for participants, analysts, and policy-makers alike.

This network of reflexive relations operates as a knowledge-intensive subdynamics of intentions, strategies, and projects that adds surplus value by reorganizing and harmonizing the political and economic structures in order to achieve a better approximation of the variety of (uncertain) goals. The issue of how much and under which conditions anyone is in control given this layer of interacting expectations specifies a research program for innovation studies.

In the case of knowledge-based innovation systems the expectation is that 'what you see is *not* what you get'! What you see, are the historical footprints of previous operations. The definition and consequently the delineation of innovation systems is knowledge-intensive. The interacting subdynamics, that is, specific operations like markets and technological innovations, are continuously reconstructed—like e-commerce on the Internet—yet differently at different places and various levels. What is considered as 'industry' and what as 'markets' cannot be taken for granted and should not be reified. Each 'system' is defined and can be redefined as research projects are further designed.

For example, 'national systems of innovation' can be more or less systemic. The extent of systemness can be studied as an empirical question (Leydesdorff and Oomes 1999). Dynamic 'system(s) of innovation' may consist of increasingly complex collaborations across national borders and among researchers and users of research from various institutional spheres (Godin and Gingras 2000). Among other things, one may expect differences among regions (Braczyk et al. 1998) and sectors (Pavitt 1984).

All these systems of reference can be specified analytically, but their systemness remains a hypothesis (Leydesdorff 2000). The Triple Helix hypothesis states that the 'systems' can be expected to remain in transition (Etzkowitz and Leydesdorff 1998). Can the observations then still provide us with an opportunity to update the expectations?

## THE STATUS OF THE OBSERVABLES

Non-linear models of innovation extend linear models by taking interactive and recursive terms into account. The non-linear terms can be expected to change the causal relations between input and output, that is, the production rules in the systems under study. The reflexivity in the discourses adds an emerging layer of learning to the evolving systems. The reflections remain structurally coupled to the actors involved as the carriers, but one subdynamics may be repressed (deselected) at the social level more than another given historical contingencies.

By changing the unit of analysis (or the unit of operation) reflexively, one is able to obtain a different perspective on the systems under study. But the latter are evolving at the same time. In terms of methodologies, this continuous change at various levels challenges our conceptual apparatus, since one is not always able to distinguish clearly whether a variable has changed ($dx/dt = f(x)$) or merely the relative value of the variable ($y = f(x)$).

Discursive analyses provide us with snapshots, while reality presents a moving picture. The analysts, however, need geometrical metaphors to render the complexity accessible to understanding and communication, and these metaphors can be stabilized by higher-order codifications, as in the case of paradigms. An understanding in terms of fluxes (that is, how the variables as well as the values change over time) calls for the use of a calculus and algorithmic simulations. The observables can then be considered as special cases that inform the expectations over time (Leydesdorff 1995).

The study of innovation systems requires this level of sophistication, since innovations can be defined only in terms of an operation which one can expect to contain both recursivity (stability) and interactivity (change). Knowledge-based innovation is therefore a cultural achievement: the innovators themselves are reflexive with respect to previous solutions. Both the innovator(s) and the innovated system(s) are expected to be changed by innovation. As noted, one is able to be both a participant and an observer, and one needs to be able to change these perspectives reflexively in order to maintain a position within the process.

Although the different narratives are mixed and confused in 'real life' events, the various models can be distinguished analytically. In his study of 'artificial life', Langton (1989) proposed to distinguish between a 'phenotypical' level of the observables and a 'genotypical' level of analytical theorizing. The 'phenotypes' remain to be explained, while the various theories compete in terms of their 'idealtypical' clarity and their usefulness in updating the respective expectations. Confusion, however, is difficult to avoid given the 'real life' pressures to jump to conclusions. The different perspectives are continuously competing, both normatively and analytically.

**EPISTEMOLOGICAL IMPLICATIONS**

The innovation systems under study can be expected to contain a complex dynamics and therefore, they do not have to be integrated nor completely differentiated. On the contrary, the delicate balance between integration and differentiation allows the system to further develop. Under selection pressure the innovation systems can be expected to perform on the edges of fractional differentiations and local integrations. Using this model of partial and temporary solutions, one can expect that the knowledge-based regime exhibits itself in terms of progressive instances and non-periodical crises.

The locally observable sequences inform us about global developments in terms of the deselected exceptions which can be replicated and built upon. The selection mechanisms, however, remain theoretical constructs. Historical case studies provide us with positive instances that enable the analyst to specify these (negative) selection mechanisms, but only reflexively. Over time, the initial inferences can perhaps be corroborated. At this end, the function of reflexive inferencing based on available and new theories moves the system forward by drawing attention to possibilities of (potentially counter-intuitive) change.

Reflexive translations can tolerate inconsistencies and differences to be resolved over time. The ensuing puzzles set a research agenda. In general, translations operate among different codifications. The exchanges at the interfaces enable us to transfer insights from one social domain to another by placing them in a different context, and thus by providing the substantive information with another meaning.

For example, in science one investigates whether a given statement is true or false, while in the economy one assesses whether one can utilize a finding (e.g., the patent) to make a profit. For the latter purpose, one does not always have to understand the underlying mechanism in great detail. However, one cannot buy the 'truth' of a statement on the market or claim it with only political power (Luhmann 1984).

When institutional differentiation (in the retention mechanism) is added to the functional differentiation in the exchange, the theoretical specification becomes one step more complex than before. The possibility of interaction increasingly adds a new dimension to the system. As noted, this expansion of the economy became urgent from 1870 onwards, when industrial R&D laboratories were installed and when patent legislation increasingly allowed for the transfer of insights from the laboratory into practice and *vice versa*. In principle, the two operations—of functional differentiation and institutional transfer—can operate as selectors upon each other, but only at certain places can resonances into coevolutions ('mutual shapings') be expected. Thus, in most places the options will not match, but, in a scattered and distributed mode, a new mode of knowledge production can increasingly be generated. The local resonances compete at a next-order systems level. In order to study this selection, however, this 'virtual' (next-order) system has first to be hypothesized and specified.

The communications can be valued systematically from different perspectives. In the case of an emerging network, each communication can have a meaning for systems other than the one in which it was generated. Since communications can be communicated, these networked systems generate internal complexities that require interface management. What competencies belong to which domains? As the organization of the interfaces and the control functions within them then also become degrees of freedom, the unambiguous attribution of a communication to either of the systems becomes increasingly difficult.

A comparison may be helpful here: translations of texts into other languages can be unambiguous because the grammars and dictionaries of the various languages are codified. Thus, if one feels uncertain, one can look up the correct translation in a dictionary or one may ask a native speaker. The situation among functional domains, however, is one in which the translators are discussing about the proper translations among themselves. When the translators happen to agree, a new codification may be more useful than the 'natural' (that is, previously given) ones.

In a complex dynamics, each system remains under reconstruction and in evolutionary competition while reorganizing complexity within its relevant environments. Since all these systems are increasingly knowledge intensive, their borders remain uncertain expectations. New codifications reconstructed by the translations may then become more functional than the originally translated ones (or not).

First, the systematic use of science in industry in the late 19$^{th}$ century raised philosophical questions about the *demarcation* between science and non-science at the interfaces (Popper 1963). This issue was solved by the so-called 'linguistic turn in the philosophy of science' during the interbellum (Rorty 1992). While previously truth had been associated with ideas, a truth-value would henceforth be attributed to statements: some statements are more likely to be true than others.

The post-modern turn has changed the situation again: the truth-value of a statement is also contextual. One has a degree of freedom to play with the centrality of concepts in terms of heuristics and puzzle solving (Simon 1969). For example, the precise definition of 'atomic weight' may differ between chemical physics and physical chemistry without necessarily creating confusion (Kuhn 1962). Concepts have meaning within discourses. Translations between discourses and reformulations within discourses can nowadays be considered as the carriers of knowledge-based developments.

This communicative turn implies neither arbitrariness in what is true or not, nor a relativistic position. The various values of a communication (including the potential truth) can only be discussed from within a discourse. These discourses are themselves developing and thus changing in terms of what is true or not. Although the discourse is uncertain in terms of its boundaries, it can also be expected to be more certain in terms of its core. Codifications structure the discourses, and translations enable us to communicate among them.

Note that the translations are asymmetrical. One cannot force validity upon a scientific discourse from a political perspective (Lecourt 1976). Analogously, the validity of a scientific statement does not guarantee its diffusion. The systems of reference have to be specified first, and then one can raise questions with respect to quality control in these various dimensions and the functions of the filters at interfaces.

For example, one can distinguish between (a) validation problems which are generated within the scientific communication system because of the differences between 'Mode 1' and 'Mode 2' research, and (b) validation problems within the knowledge produced in the applicational contexts of 'Mode 2' (Fujigaki and Leydesdorff 2000). New mechanisms of quality control can sometimes be expected to emerge (for example, at the Internet). Thus, the new mode of the production of scientific knowledge can empirically be operationalized in terms of epistemological domains, that is, with reference to the validity of scientific knowledge in various contexts, yet without harming the standard conventions of scientific soundness in terms of true and false statements (Leydesdorff 2001b).

## METHODOLOGICAL CONSEQUENCES

Since the reflexive selection mechanisms of cultural evolution cannot be identified by unmediated observation, the neo-evolutionary analysis has to begin with the specification of a hypothesis. 'What' does one expect to be communicated and why? The study of how this communication is institutionally arranged (and therefore measurable, in principle) is then a question of empirical design.

From this perspective, the observable arrangements have a status beyond merely providing the analyst with one or another, as yet unreflexive starting point for the narrative. The data can be used for informing *ex ante*—and sometimes testing *ex post*—the theoretical expectations. Which layer operated with which function, why and in which instances? By raising first the substantive question of 'what is communicated?'—e.g., economic expectations (in terms of profit and growth), theoretical expectations or assessments of what can be realized given institutional and geographic constraints—the focus is firmly set on the specification of the media of communication. How are these media of exchange related and converted into one another? Why are some processes more mutually attractive than others, and under what conditions can the exchanges among them be sustained?

Although the discourses are able to interact at the interfaces, the frequency of such external interactions is (at least initially) lower than the frequency of interactions within each helix (Simon 1969). The helices communicate recursively over time first in terms of their own respective codes. Reflexively, they can sometimes take the role of each other, yet only to a certain extent. Over time and with the availability of ICT, this relation may begin to change (Cowan and Foray 1997). One can expect that the balance between spatial and virtual relations is contingent upon the availability of the exchange media and their respective codifications. The maintenance of codification may be costly, but it enables the systems to suppress noise in the communication.

Inter-human communication remains failure prone. Quality control of communication is crucial for developing the knowledge base of the system, but it remains a 'counter-factual' expectation. One experiences mainly the misfits between the various modes of communication. The selective operations can be specified with hindsight, that is, as hypotheses on the basis of theoretical inferences that recombine on the basis of insights in various disciplines.

Despite this 'virtuality' of the knowledge-based operation at the level of the overlay (Giddens 1984), the socio-economic system—that is potentially innovated by the operation—is not 'on the fly': it remains also grounded in a culture which has to be reproduced in terms of renewing the social systems of coordination. However, the retention mechanisms are no longer given; the institutonal layer is increasingly 'on the move.' It can be reconstructed as the system is deconstructed, that is, as one of its subdynamics.

## CONCLUSION

The emerging system rests like a hyper-network on the networks on which it builds (such as the disciplines, the industries, and the national governments), but the knowledge-economy transforms 'the ship while a storm is raging on the open sea' (Neurath 1933). As the technological culture provides options for recombination, the boundaries of the carrying communities can also be reconstituted. The price of changing the communal base may be felt as a loss of traditional identities, alienation, or as a concern with the sustainability of the reconstruction. However, 'creative destruction' entails the option of increasing development. In this sense, knowledge-based innovation can be turned into 'a celebration of community' because the new mode of production is based on reaching across previously established boundaries (Hayward 1998).

I have argued that the evolutionary perspective in economics can be complemented with a turn towards reflexivity in sociology in order to obtain a richer understanding of how the overlay of communications in university-industry-government relations reshapes the systems of innovations that are currently subjects of debate, policy-making, and scientific study. Without this reflexive turn, evolutionary studies tend to reify systems like 'national systems of innovation,' 'regional innovation systems,' or sectors of the economy. Both participants and analysts, however, can always redefine systems. Furthermore, the participants and the analysts use their system's definition in studying and changing the system.

Whereas a coevolution between, for example, markets and technologies can be entrained along a trajectory, a triple helix of university-industry-government relations can no longer be expected to stabilize. The global regime may contain competing trajectories. The analyst can hypothesize and then also try to measure whether a technological development (e.g., aircraft) is in its trajectory phase or in the regime phase (e.g., Frenken and Leydesdorff 2000). In the regime phase, knowledge-based innovation policies should be aimed at influencing the political economy of a technology at the regime level and not merely at addressing the trajectory of its further improvements.

**REFERENCES**

Allen, Peter M. 1994. "Evolutionary Complex Systems: Models of Technology Change." In *Evolutionary Economics and Chaos Theory: New Directions in Technology Studies*, edited by Leydesdorff & Van den Besselaar. London: Pinter.
Anderson, E. S. 1994. *Evolutionary Economics: Post-Schumpterian Contributions*. London: Pinter.
Arthur, W. Brian. 1988. "Competing Technologies." In "Technological Paradigms and Technological Trajectories: A Suggested Interpretation of the Determinants and Directions of Technical Change," edited by Dosi et al. *Research Policy* 11, 147–162.
Arthur, W. Brian 1989. "Competing Technologies, Increasing Returns, and Lock-In by Historical Events." *Economic Journal* 99, 116–131.
Arthur, W. Brian. 1994. *Increasing Returns and Path Dependence in the Economy*. Ann Arbor: University of Michigan Press.
Barnes, Barry and David Edge 1982. *Science in Context*. Cambridge, MA: MIT Press.
Bijker, Wiebe, Thomas P. Hughes, and Trevor Pinch, eds. 1987. *The Social Construction of Technological Systems*. Cambridge, MA: MIT Press.
Braczyk, Hans-Joachim, Philip Cooke, and Martin Heidenreich, eds. 1998. *Regional Innovation Systems*. London/Bristol PA: University College of London Press.
Braverman, Harry. 1974. *Labor and Monopoly Capital*. New York/London: Monthly Review Press.